\documentclass[useAMS]{mn2e}
\usepackage{epsfig}
\usepackage{graphicx}
\usepackage{epsf}
\usepackage{amsmath, amssymb}
\newcommand{\plotone}[1]{\includegraphics[width=84mm]{#1}}

\title[Eccentricity Growth Rates of Tidally Distorted Discs]
{ Eccentricity Growth Rates of Tidally Distorted Discs}

\author[Stephen H. Lubow]
{ Stephen H. Lubow$^{1,2}$
\\
$^1$STScI, 3700 San Martin Drive, Baltimore, MD 21218, USA\\ 
$^2$Institute of Astronomy, Madingley Road, Cambridge, CB3 0HA, UK\\}

\begin{document}
\maketitle

\begin{abstract}

We consider 
discs that orbit a central object and are tidally perturbed by a circular orbit companion.
Such discs are sometimes subject 
to an eccentric instability due to the effects of certain resonances. 
Eccentric instabilities may be present in planetary rings perturbed by satellites,
protostellar discs perturbed by planets, and discs in binary star systems.
Although the basic mechanism for eccentric instability is well understood,
the detailed response of a gaseous disc to such an instability is not understood.
We apply a linear eccentricity evolution
equation developed by Goodchild and Ogilvie.
We explore how the eccentricity is distributed in such a disc and how
the distribution in turn affects the instability growth rate for a range of disc properties.
We identify a disc mode, termed the superhump mode, that is likely
at work in the superhump binary star case. The mode results from the excitation
of the fundamental  free precession
mode. We determine an analytic expression
for the fundamental free mode precession rate that is applicable to a sufficiently cool disc. 
Depending on the disc sound speed and disc edge location, other eccentric modes can
grow faster than the superhump mode and 
dominate. 
 
\end{abstract}



\section{INTRODUCTION}

Consider a gaseous disc that orbits about a central object. If the disc 
is tidally perturbed by a circular orbit companion, but the disc is otherwise circular, then its response at 
certain (Lindblad) resonance
locations can be quite strong and result in the launching of waves that extract
energy and angular momentum from the companion (Goldreich \& Tremaine 1979).
If the perturber were to be removed, the disc would settle back to a circular state, with its mass
having been rearranged from its initial state by the resonant torques.  
For discs that are slightly eccentric, additional resonances arise. Some of these resonances
have the property that they can cause the disc eccentricity to grow exponentially
in time, even though the perturber is on a circular orbit. In this case, if the perturber were to be removed, the disc would be in an eccentric state.
The process by which eccentricity growth occurs in a fluid disc can be understood through a mode-coupling
mechanism (Lubow 1991a; hereafter L91). The instability can also be understood in terms 
of particle dynamics with dissipation (Borderies, Goldreich, \& Tremaine 1983, Ogilvie 2007).
Such instabilities may explain the eccentricities of planetary rings perturbed by
satellites (Borderies et al 1983), 
the eccentricities of circumstellar discs perturbed by planets or brown dwarfs (Papaloizou, Nelson, \& Masset 2001,
Kley \& Dirksen 2005, D'Angelo, Lubow, \& Bate 2006),
and 
eccentricities in discs of catalysmic binaries
 (Whitehurst 1988, Lubow 1991b,  Osaki 1996),  and X-ray binaries (Haswell et al. 2001, Neil, Balyn, \& Cobb 2007).

In this paper, we will mainly consider the superhump binary case which has been 
 studied, both observationally and theoretically. We will generalise the results somewhat
 to cover cases involving warmer (larger dimensionless disc thickness) protostellar discs. 
 The initial observational evidence for eccentric
 discs came from an analysis of a drifting feature, the superhump, in the light curves
 of extreme mass ratio cataclysmic binaries  that undergo unusually strong outbursts called 
 superoutbursts (Warner 1975).
 The period of the superhump feature is slightly longer than the binary orbit period.
 The drifting feature in the light curve was attributed to an eccentric prograde precessing
 disc (Vogt 1982, Osaki 1985). The prograde precession is due to the gravitational effects of the companion 
 that causes departures of disc streamlines from closed Keplerian ellipses. 
 Weaker retrograde effects occur due
 to gas pressure (Lubow 1992, Murray 2000).  Simulations by Whitehurst (1988) revealed an eccentric instability
 that can be understood through the effects of the 3:1 eccentric Lindblad resonance (L91).
 Many observational studies have supported the idea that the disc is eccentric 
 (e.g., Osaki 2003, Patterson 2005).
 
 Several particle based simulations (SPH or something similar) of discs in binaries revealed
 properties of the instability and the disc precession (see, e.g., Smith et al 2007). For many years, 
 simulations carried out by grid-based codes were unable to
find this instability and cast some doubt on the eccentric disc model (e.g., Heemskerk 1994, Stehle 1999).
However, Kley, Papaloizou, \& Ogilvie (2008) 
recently reported finding this instability with a grid-based code and showed
that its properties largely agree with theoretical expectations of the mode-coupling model.
However, some properties of the instability revealed 
by these simulations were not predicted by the L91 model, such as
the dependence of the growth rate on the disc sound speed.

The L91 model determined the eccentricity growth rate for a ring of gas having uniform
eccentricity. It was generalised to a full disc, again assuming the eccentricity is uniformly
distributed. However, it is by no means obvious that the disc eccentricity would always
be uniform. The distribution of eccentricity has an important influence on the
eccentricity growth rate. The resonance provides a certain rate of
angular momentum and energy input, resulting in an eccentricity growth.
But if this  angular momentum and energy input is shared over a broad eccentric
region of the disc, then we expect the eccentricity growth rate to be smaller than
if the region is narrow. In fact, the L91 model predicts that the
growth rate to vary as roughly the inverse of the radial width of the eccentric region.

Important progress in understanding the disc eccentricity distribution and its effects
on the growth rate was made by Goodchild \& Ogilvie (2006; hereafter GO06). They derived differential
equations for the radial distribution and growth rate of eccentricity, subject to the eccentric
 instability caused by the resonance. Their results for superhump binaries, involving the 3:1
 resonance, indicated that the
 disc eccentricity growth
rate was lower than expected from simulations and observations.
The reason for the small growth rate  is that the mode they analysed has a 
low amplitude near the resonance. Consequently, the resonance was ineffective
in providing eccentricity growth. We therefore have a puzzle that a more complete
modeling of the disc eccentricity evolution resulted in growth rates that are lower than expected.
In this paper we re-examine the nature of the eccentricity evolution
of a disc, using the GO06 equations, in an attempt to resolve this issue. In the
process, we determine a more strongly excited mode that we call the superhump mode.
We analyse its properties under various conditions. We also determine other modes
than can dominate, under somewhat different disc conditions than are typically expected
in superhump binaries.

The outline of the paper is as follows. In Section 2, we 
describe the eccentricity evolution equations. Section 3
specifies the standard model parameters used in the calculations.
Section 4 discusses the methods to solve the eccentricity evolution equations.
The results of the calculations for a variety of system
parameters are described in Section 5. Section 6 contains the summary.

\section{ECCENTRICITY EVOLUTION EQUATIONS}

Consider a two-dimensional disc that orbits about a primary object which
is located at the origin of a cylindrical coordinate system
 $(r, \phi)$.  
 The low-mass secondary is in a circular
 orbit of semi-major axis $d$ about the central object.
We apply the 
linearised eccentricity equations of GO06,
expressed in the following form
\begin{align}
i \partial_r ( a(r) \partial_r E(r,t) ) + i b(r) E(r,t) +  J(r) s(r)  E(r,t) \nonumber & \\ = J(r) \partial_t E(r,t),
\label{E_eq}
\end{align}
where 
$E(r,t) = e(r,t) \exp{(i \varpi(r,t))}$ is the complex eccentricity, for real
eccentricity $e$ and periapse angle $\varpi$.
This two dimensional approximation is valid under the assumption of vertical (perpendicular to the disc plane)
hydrostatic equilibrium for a thin disc. The reduction to two dimension results from
a vertically integration of three dimensional equations.
We ignore the effects of viscosity on the eccentricity. 
For the case of no perturbing body, the nonlinear evolution of an eccentric, three dimensional, viscous disc
was analysed by Ogilvie (2001).

$E$ is related to the linear perturbations from the axisymmetric circular
velocity. For the velocity expressed cylindrical coordinates as
$(u' (r,t) \exp{(-i \phi)}, v'(r,t)  \exp{(-i \phi)})$ we have that
\begin{equation}
u'(r,t) = i r \Omega(r) E(r,t)
\label{u}
\end{equation}
and
\begin{equation}
v'(r,t) = \frac{1}{2} r \Omega(r) E(r,t),
\label{v}
\end{equation} 
where $\Omega(r)$ is the Keplerian orbital frequency 
about the primary of mass $M_{p}$ given by
\begin{equation}
\Omega(r) = \sqrt{\frac{G M_p}{r^3}}.
\end{equation}
Quantity  $J(r)$ is the disc angular momentum per unit radius divided by $\pi$ and is given by
\begin{equation}
J(r) = 2 r^3 \Omega(r) \Sigma(r).
\end{equation}
Functions $a(r)$ and $b(r)$ are given by
\begin{eqnarray}
a(r) &=& \gamma P(r) r^3, \\
b(r) &=&  \frac{d P}{dr} r^2 +  J(r)  \dot{\varpi}_{\rm g},
\label{ab}
\end{eqnarray}
where
$P(r)$ is the two-dimensional (vertically integrated) disc pressure, $\Sigma(r)$ is the disc surface density, and
 $\dot{\varpi}_g$ is the gravitational precession rate 
 of a free particle on an eccentric orbit which is given by
\begin{equation}
\dot{\varpi}_g(r) = \frac{1}{4} q  \left( \frac{r}{d} \right)^2 \, \Omega(r) \, b^{(1)}_{3/2} \left( \frac{r}{d} \right).
\label{prec_g}
\end{equation}
The terms involving quantities $a$ and $b$ describe the 
eccentricity propagation and precession, respectively.
Quantity $\gamma$ is the gas adiabatic index.
Mass ratio $q$ is the mass of the secondary (perturbing) object divided by the mass of the
primary (central) object, and $b^{(1)}_{3/2}$ is the Laplace coefficient for the $m=1$ tidal potential component associated with the perturbing object.
Real function $s(r)$ is the eccentricity growth rate contribution from the eccentric instability. It is largest
in a region containing the resonance that
drives the instability. In the case of superhump binaries,
this is the 3:1 resonance.

The surface density $\Sigma(r)$ is taken to vary smoothly from 
inner radius  $r_{\rm i}$ to outer radius $r_{\rm o}$ and to be zero 
outside this range. 
Functions $a(r)$ and $b(r)$ vary smoothly
in radius, while function $s(r)$ is localised near the resonance.
For the eccentricity injection rate caused by a resonance, we generally adopt a gaussian form
\begin{equation}
s(r) = \frac{ \chi \, \exp{[-(r-r_{\rm res})^2/w_{\rm res}^2] }}{\sqrt{2 \pi} w_{\rm res}},
\label{s}
\end{equation}
where $r_{\rm res}$ is the radius of the resonance
that excites eccentricity, $w_{res}$ is the resonance width, and $\chi$
is a measure of the resonance strength.

We consider the boundary conditions at the disc inner and outer edges located
at radii $r_{\rm i}$ and $r_{\rm o}$, respectively.
Following GO06, we adopt the outer boundary condition 
\begin{equation}
\partial_r E(r_{\rm o},t) = 0.
\label{oBC}
\end{equation}
For a Keplerian disc, the divergence of the velocity is proportional to $\partial_r E$.
Consequently, this boundary condition is equivalent to requiring that the
Lagrangian density perturbation near the disc outer edge vanishes.

 The inner boundary condition is taken to be
\begin{equation}
E(r_{\rm i},t) = 0.
\label{iBC}
\end{equation}
This condition could be imposed by a hard wall inner circular boundary due to a central object.
We discuss the inner boundary condition
further in Section \ref{sec:iBC}.
 
\section{STANDARD MODEL}
\label{sec:std}

We start with a simple disc model that we define here. We later explore certain
variations from this model.
We adopt disc models in which $\gamma = 5/3$, $T \propto r^{-3/4}$,
and $\Sigma \propto r^{-3/4}$. The density form corresponds to a steady state
alpha disc with a constant value of alpha. In this model, the disc
thickness-to-radius ratio is given by
\begin{equation}
H/r = h (r/r_{\rm res})^{1/8},
\end{equation}
where $h = H/r$ evaluated at $r=r_{\rm res}$.  We do not consider any tapering
of the disc density near the the outer edge. Some tapering is needed over a distance $\sim H$
to prevent pressure forces from causing the disc to become Rayleigh unstable, i.e.,
in order that the epicyclic frequency be real. We do not include tapering, in order to
keep the standard model as simple as possible.
 In this paper, we consider
discs with various $h$ values, but take the $r-$dependence of $T$ and $\Sigma$ to be fixed.

We take the unit of length to be the binary separation $d$. 
 We consider the effects the eccentricity instability at the 3:1 eccentric Lindblad resonance.
 For $q$ small, the resonant radius is approximately given by
 \begin{equation}
 r_{\rm res}= 3^{-2/3} (1+q)^{-1/3}. 
 \label{rres}
 \end{equation}
 The 3:1 resonance strength $\chi$ in equation (\ref{s}) is given by
 \begin{equation}
 \chi = 2.08 \, q^2 \Omega_{\rm b} r_{\rm res},
 \label{chi}
 \end{equation}
 where $ \Omega_{\rm b}$ is the binary orbital frequency (L91).
 This expression was independently confirmed by the particle model
 of Ogilvie (2007).
The resonance width $w_{\rm res}$ depends on the disc sound speed and we take
\begin{equation}
w_{\rm res} = h^{2/3}\, r_{\rm res}
\label{w}
\end{equation}
(see Meyer-Vernet \& Sicardy 1987).
We consider variations from this value of the width in Section \ref{sec:grw}.

We adopt a mass ratio $q=0.1$
as a standard value.
With this mass ratio, the disc outer radius determined by Paczynski's (1977) particle orbit intersection
method is equal to 0.46. The disc radius given by the 
prescription suggested in Whitehurst \& King (1991) evaluates to 0.53, which is
 0.9 times the Roche lobe radius.
We adopt a fiducial value of 0.5. The disc inner radius is chosen as 0.01 for
numerical convenience to avoid the singularity at the disc center. 
Smaller inner disc radius values have little impact
on the results presented here, as is discussed in Section \ref{sec:iBC}.

In summary, we adopt a set of standard parameter values to model superhump binaries
and later consider variations from some of these values. The standard model
has $q=0.1$, $h=0.02$, $r_{\rm i} =0.01$, $r_{\rm o}=0.5$,  $r_{\rm res}=0.466$, $\gamma = 5/3$, $T \propto r^{-3/4}$,
and $\Sigma \propto r^{-3/4}$.

\section{METHODS OF SOLUTION}
\label{sec:mos}
Equation (\ref{E_eq}) and
boundary conditions (\ref{oBC}) and  (\ref{iBC}) are homogeneous and therefore
permit the specification of the value of $E(r,0)$ at some selected radius $r$.
For the standard model we adopt
a normalization
\begin{equation}
E(r_{\rm o},0) = 1.
\label{norm}
\end{equation}
Of course, we are not suggesting that the actual eccentricity is unity.
But, since the equations are linear, the solutions can be scaled linearly
to any desired normalization.

Equation  (\ref{E_eq}) can be solved by searching for eigenmodes in which
\begin{equation}
E(r,t) = E(r) \exp{(i \omega t)}
\label{Eom}
\end{equation}
and complex frequency $\omega$ is the eigenvalue.
The function $E(r)$ and eigenfrequency $\omega$ can be determined by shooting methods
that satisfy the boundary and normalization conditions. 
Equation  (\ref{E_eq}) is second order in space and
can be integrated inward in radius starting at $r=r_{\rm o}$ by using the two conditions,
equations (\ref{oBC})  and (\ref{norm}), and an assumed value for $\omega$.
Complex frequency $\omega$ is adjusted until the inner boundary condition, 
equation (\ref{iBC}), is satisfied by using Newton's method.
Another approach would be to discretise  equation (\ref{E_eq}) in radius and solve
the equations represented by a sparse matrix.

Such approaches have the drawback that there are many modes in the system and one
must search for the physically relevant one that is the fastest growing.
The shooting method is particularly prone to missing the fastest growing mode.
The reason is that the guessed value for the eigenfrequency must be close
to the value for the fastest growing mode. Otherwise, the method will typically
converge on another eigenfrequency of the system. 

We have instead opted to find the fastest growing mode by integrating equation
 (\ref{E_eq}) in space and time as a PDE. We apply the following initial condition
 \begin{equation}
 E(r,0)=\cos{\left[ \frac{\pi}{2} \left(\frac{r_{\rm o}-r}{r_{\rm o}-r_{\rm i}} \right) \right]}
 \label{ic}
 \end{equation}
 for $r_{\rm i} \le r \le r_{\rm o}$.
 This initial eccentricity satisfies boundary and normalization conditions (\ref{oBC}),  (\ref{iBC}), and (\ref{norm}).
 Over time $E(r,t)/E( r_{\rm o},t)$ settles to an eigenfunction in which $\partial_t E/E$ is constant
 in space and is the complex eigenfrequency. This approach is computationally much slower than solving
 the eigenvalue problem with the shooting method, if one has a good initial guess for the eigenfrequency.
  We apply the eigenfrequency obtained from the time integration to the shooting method described above
 to verify its existence. We then smoothly vary other parameters such as $q$ and
 $h$ to rapidly obtain many other solutions by  iteratively applying the eigenfrequency from the last
 obtained solution as the initial
 guess of the eigenfrequency for the next nearby set of parameters. In this way,
 we obtain a set of solutions, along a continuous branch that began with a fastest growing mode.
 There is no guarantee, however, that all solutions in this branch are the fastest growing
 for their set of parameters. Sufficiently nearby solutions are typically found to also be the fastest growing.
 However, as we will see,  multiple branches occur that contain the fastest
 growing modes in different parameter regimes.
 
 To integrate equation (\ref{E_eq}), we initially wrote a C program
 to implement the Crank-Nicholson method (e.g., Watanabe \& Tsukada 2000). This method is unconditionally
 stable and fast, but produced some unwanted oscillations at high resolution, possibly
 as a consequence of Gudunov's Theorem (Wesseling 2001). Instead, we opted for the method of lines
 in Mathematica.
 With this method, the eccentricity is discretised in radius, producing a set of eccentricities
 $E_{\rm i}(t)$, one at each radial grid point ${\rm i}$. Equation (\ref{E_eq}) becomes to a set of coupled
 ODEs in time that are integrated by high order schemes. This method produced high resolution
 results.
 
 Based on GO96, we express the  growth rate for the standard model 
 as
 \begin{equation}
 -Im(\omega) = \frac{ \int r^{3/4}  s(r) e^2(r) dr }{\int r^{3/4} e^2(r) dr} \simeq \frac{\chi}{max(w_e,w_{\rm res})}
 \left( \frac{e(r_{\rm res})}{e_{max}} \right)^2 ,
 \label{gr2}
 \end{equation}
 where the integrals are taken over the disc from $r=r_{\rm i}$ to $r=r_{\rm o}$,
 $e_{max}$ is the maximum eccentricity in the disc, and $w_e$ is the radial width
 of the eccentricity distribution. This equation demonstrates the dependence of the growth
 rate on the properties of the eccentricity distribution.

\section{RESULTS}
\subsection{Superhump Mode}

\begin{figure*}
\centering%
\resizebox{\linewidth}{!}{%
\includegraphics[height=30pt]{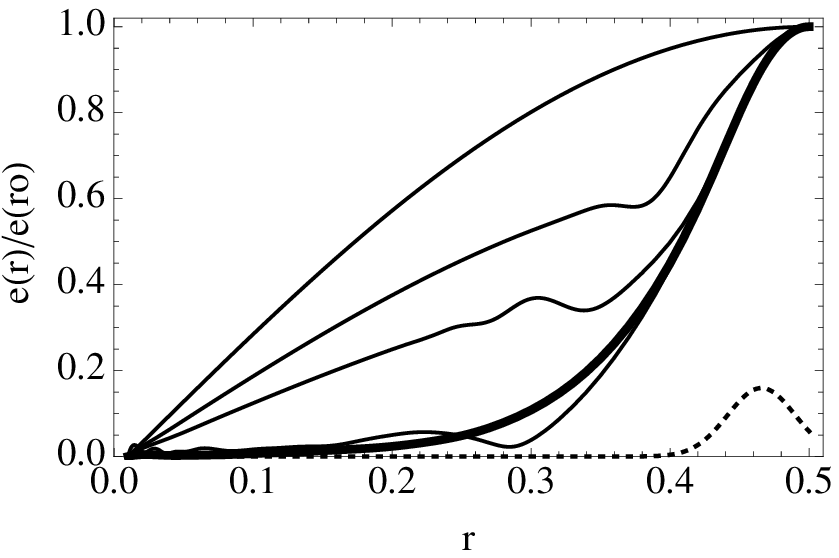}%
\hspace{0.1in}
\includegraphics[height=30pt]{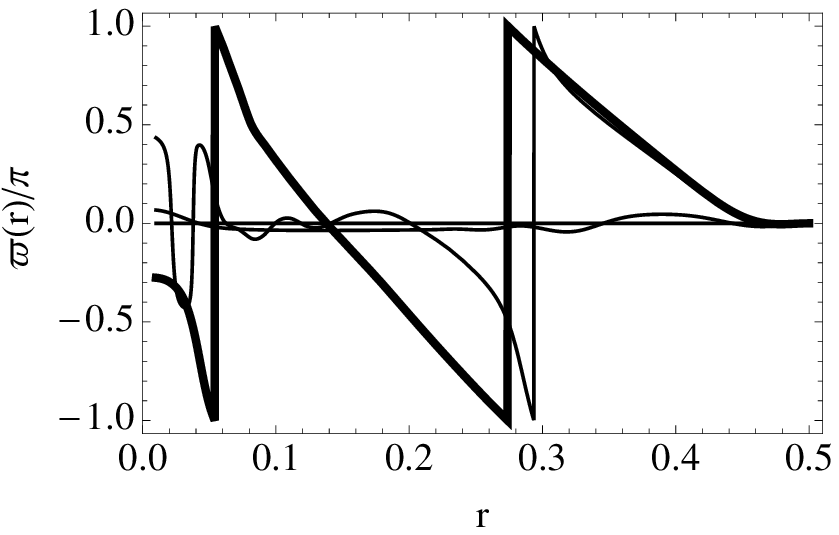}
}
\caption{Evolution of eccentricity for the standard model ($q=0.1, h=0.02, r_{\rm o}=0.5$). The initial
eccentricity distribution is of a cosine form, equation (\ref{ic}). 
Radius $r$ is in units of the binary separation.
\textit{Left panel:}   $e(r)$ at various times, normalised
by its value at the disc outer edge.
The curves montonically drop in time at  $r=0.2$,
corresponding to times of 0, 1, 2, 8 and 16 (thick line) binary orbits. The dotted
curve is the 3:1 resonance instability factor $s(r)$ of
equation (\ref{E_eq}) in units of $\Omega_{\rm b}$.
\textit{Right panel:} Phase angle  (periapse angle) $\varpi(r)$ 
of the eccentricity in radians divided by $\pi$ as a function 
of radius at various times. The curves monotonically increase in time at $r=0.4$, corresponding
to times of 0 (flat line of zero phase), 2, 8, and 16 (thick line) binary orbits.  
         }
\label{fig:mols}
\end{figure*}

The method of lines solution to equation (\ref{E_eq}) for the standard model, subject to initial
condition (\ref{ic}) and boundary conditions (\ref{oBC}) and  (\ref{iBC}), is plotted in Fig.~\ref{fig:mols}.  After a time of about
16 binary orbital periods, the eccentricity distribution settles into a mode.
The values of the normalised eccentricity $e(r,t)/e(r_{\rm o},t)$ and the
phase angle  $\varpi(r,t)$ become nearly time-independent.
The eccentricity growth rate $\partial_t e(r,t)\,/e(r,t)$ and precession rate 
$\partial_t  \varpi(r,t)$ become independent 
of radius. 

The eccentricity growth rate for the standard model is determined to be
\begin{equation} 
\frac{\partial_t e(r,t)}{e(r,t)} = -Im(\omega) \simeq 0.10 \Omega_{\rm b}, 
\label{grs}
\end{equation}
where $\omega = \partial_t E(r,t)/E(r,t)$. 

This rapid growth rate is  a consequence of the strong overlap between the eccentricity 
distribution in the mode with the eccentricity driving by the resonance and the narrowness of
the eccentricity distribution.
From the figure, we estimate $w_e \simeq 0.1$ and 
$e(r_{\rm res}) \simeq e_{max}$. 
Applying the  parameters
for the standard model ($q=0.1$ and $r_{\rm res} \simeq 0.466$), we obtain 
an estimated growth rate from equation (\ref{gr2}) that is $- Im(\omega) \simeq 0.1  \Omega_{\rm b},$ in agreement with the rate obtained from
the time-evolution solution, equation (\ref{grs}). Therefore, we find that the eccentricity
is robustly excited in the outer parts of the disc. This calculation is telling
us something that  L91 did not, the effects of the eccentricity distribution on the growth rate. 
As we will see, the width  of the eccentricity distribution $w_e$  is a function of the system parameters
and the growth rate itself.

The mode shown in the figure is termed the superhump mode. 
To understand its properties, we associate a timescale with each coefficient on the left-hand side of equation (\ref{E_eq}).
The pressure, precession, and resonance instability timescales are estimated
as $w_e^2 J/a$, $J/b$ and $r_{\rm o}/\chi$, respectively. For the standard model, they
crudely evaluate near the resonance to  $10^3 w_e^2/(r_{\rm o}^2 \Omega_b), 
\, 20/\Omega_b,$ and $50/\Omega_b$, respectively.
The effects of pressure act to spread the eccentricity distribution 
and balance against the effects of precession and resonance instability that will be shown to confine the eccentricity.
 The eccentricity distribution width $w_e$ must be substantially smaller than $r_{\rm o}$, so that the pressure
 timescale is
comparable to the precession and instability timescales.
 Therefore, the eccentricity should be radially confined to a region smaller than the disc radius. 

Notice that the phase of the mode decreases with radius, indicating
that the mode is a trailing waveform. The radial wavelength
is long, about equal to 0.3.
Recall that the disc scale height is only about $2\%$ of the disc radius or 0.01.

As outlined in Section \ref{sec:mos}, we apply the complex eigenfrequency obtained
for the standard model to solve the mode equation by the shooting method. We then smoothly vary parameters
to determine how the growth rate and modal structure varies as a function of various parameters, as will
be described in the subsequent subsections.

\subsection{Variations in $q$}

We consider how the growth rate varies with binary mass ratio, $q$.
As $q$ varies, the radius of the 3:1 resonance varies slightly, according to
equation (\ref{w}). But since the resonance lies close to the disc outer edge,
varying $q$ alone would introduce an additional effect in changing the clearance
between resonance and disc edge that we denote as 
\begin{equation}
\Delta r_{\rm c} = r_{\rm o}-r_{\rm res}.
\end{equation}
We analyse the effects of changing the clearance $\Delta r_{\rm c}$ separately.
Consequently, when varying $q$ we also vary $r_{\rm o}$ so as to maintain
a constant value $\Delta r_{\rm c} = 0.5 - 0.466 = 0.034$. This means that $r_{\rm o}$
varies slightly with $q$.  

\begin{figure*}
\centering%
\resizebox{\linewidth}{!}{%
\includegraphics[height=32pt]{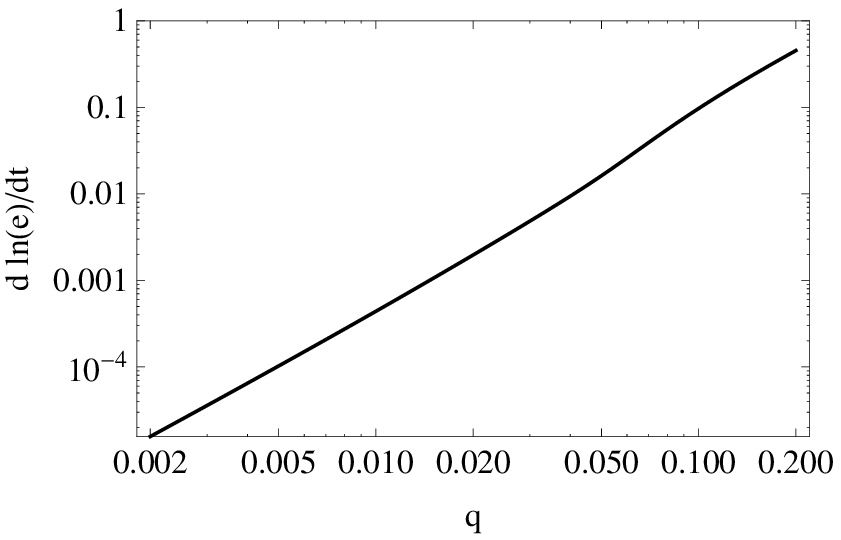}%
\hspace{0.1in}
\includegraphics[height=30pt]{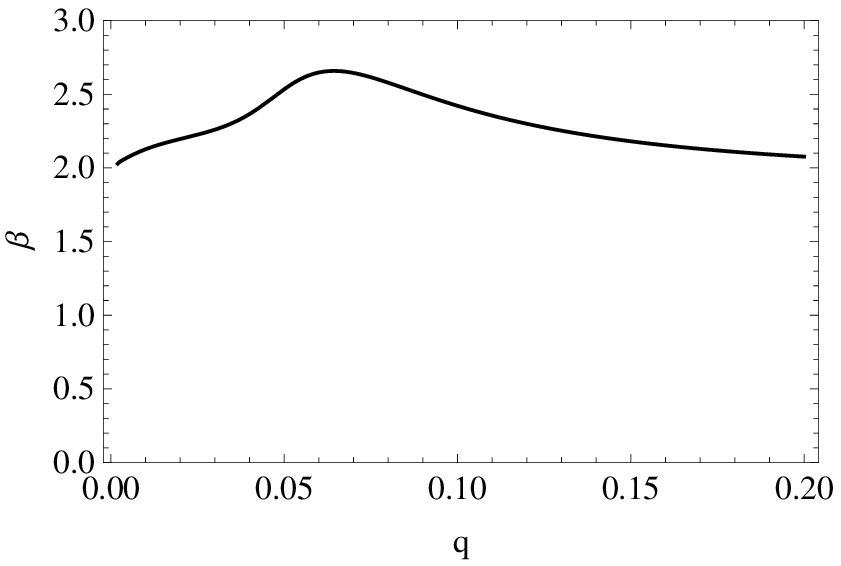}
}
\caption{
\textit{Left panel:} Eccentricity growth rate in units of $\Omega_{\rm b}$. plotted against
binary mass ratio $q$ for the standard model. The disc outer edge varies slightly with $q$ from the standard value of 0.5, so that
the distance from the resonance to the disc outer edge is maintained at a constant value.
\textit{Right panel:} Plot of $\beta$, defined by $d \ln{(e)}/dt \propto q^\beta$ as a function of $q$
based on the results of the left panel.
         }
\label{fig:grq}
\end{figure*}

\begin{figure}
\includegraphics[width=84mm]{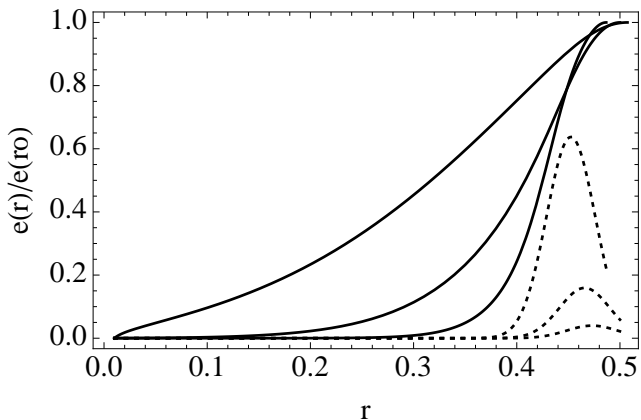}
\caption{Eccentricity distributions for cases with standard parameters, but having
different values of binary mass ratio $q$. The $q$ values from the  highest to lowest solid curves are 
0.05, 0.1, and 0.2, respectively.  The disc outer edge varies slightly with $q$ from the standard value of 0.5, so that
the distance from the resonance to the disc outer edge is maintained at a constant value.
The dotted curves are the  3:1 resonance instability factors $s(r)$ of
equation (\ref{E_eq}) in units of $\Omega_{\rm b}$ for the same set of $q$ values, where the
peak values increase with increasing $q$. The eccentricity distributions narrow
with increasing $q$, since the increasing growth rates cause the eccentricity to be more
concentrated near the resonance. 
         }
\label{fig:e-q1}
\end{figure}

We carried out a sequence of solutions to the modal equations using the shooting method for
different binary mass ratios $q$. 
The results plotted in Fig.~\ref{fig:grq} show that the growth rate varies
somewhat faster than $q^2$. A quadratic dependence of the growth rate on $q$ follows from the $\chi$
term (see equation (\ref{chi})) in equation (\ref{gr2}). The somewhat more rapid dependence 
is a consequence of additional mode confinement with increasing $q$ (see Fig.~\ref{fig:e-q1}).
The confinement is a result of a competition between the pressure  that spreads
the distribution with the eccentricity growth that narrows the distribution to a region near the resonance.

\subsection{Free eccentric mode}

\begin{figure}
\includegraphics[width=84mm]{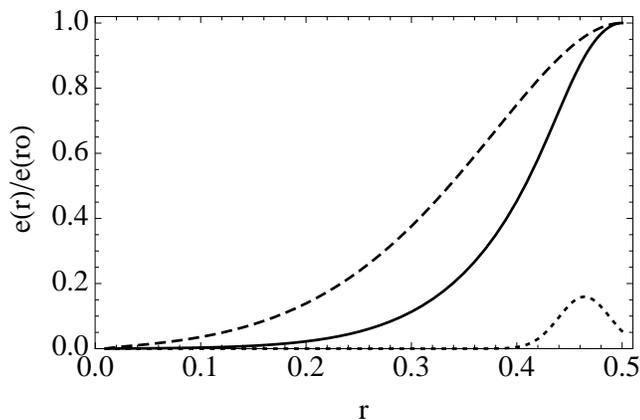}
\caption{Eccentricity distribution for a system with standard parameters (solid line) 
and a system with zero resonance strength (dashed line), obtained through a sequence of models
in which the resonance strength of the standard parameter case is smoothly lowered to zero.
 The dotted
line is the 3:1 resonance instability factor $s(r)$ of
equation (\ref{E_eq}) in units of $\Omega_{\rm b}$ for the standard case
(solid line). The 3:1 resonance in the outer parts of the disc acts to further confine the mode near the disc outer
edge. The dashed curve is the fundamental free eccentric mode of the disc. It is that mode which is excited
by the eccentricity instability to become the superhump mode.
         }
\label{fig:freem}
\end{figure}

To determine the free eccentric mode corresponding to the superhump mode,
we regarded $q$ as fixed and constructed a sequence of models starting with the superhump mode  
and smoothly reduced the resonance strength $\chi$ (or $s(r)$) to zero.
In that limit, we obtain a mode structure that is plotted in Fig.~\ref{fig:freem},
along with the mode structure in the standard model.
Both modes are computed with the same disc outer radius equal to 0.5.
The $\chi=0$ mode corresponds to the fundamental (lowest order) free eccentric mode. Therefore,
we see that the superhump mode results from the excitation of the free eccentric mode.

The additional mode confinement in the case of the growing mode can again  be understood in
terms of a competition between the radial pressure-induced propagation of eccentricity that acts to spread
the mode and its
growth that acts to localise the mode close to the 3:1 resonance that resides near the disc outer edge (the dotted
curve in the figure).  The confinement in the free mode case is shown below to be due to the effects of disc precession.

A previous  estimate of eccentric disc free
precession rate, based on the WKB theory of density waves, 
suggested
\begin{equation}
\omega = \dot{\varpi}_{\rm g}(r_{\rm o})  - \frac{ k^2 H^2}{2 } \Omega(r_{\rm o}),
\end{equation}
where $k$ is the radial wavenumber of the eccentric mode
(Lubow 1992).
This result demonstrates that pressure causes retrograde precession (by the second term on
the right-hand side), opposite
to the prograde precession due to the companion (the first term on the right-hand side).
However, the magnitude of $k$  was not determined.

We analyse the fundamental  free precession mode for a cool disc by means of
equation (\ref{E_eq}), together with equation (\ref{Eom}) and  the condition $s(r)=0$.
For a cool disc, we expect that the mode is confined to a small
radial region near the  disc outer edge and so $ |d E/dr| >> |E|/r$.
In addition, we expect that $\omega \simeq  \dot{\varpi}_{\rm g}(r_{\rm o}).$
We apply these approximations to obtain
\begin{equation}
a(r_{\rm o}) \frac{d^2 E}{d r^2} +J(r_{\rm o}) (\dot{\varpi}_{\rm g}(r) - \omega) E(r) = 0.
\label{Ey}
\end{equation}
We expand $\dot{\varpi}_{\rm g}(r)$ in a Taylor series to linear order about $r_{\rm o}$ and then have
\begin{equation}
r_{\rm o}^2 \, \, \frac{d^2 E}{d r^2} + (c_0 + c_1 \, \Delta r )E(r) = 0,
\label{Ez}
\end{equation}
where $\Delta r = r - r_{\rm o}$,
\begin{equation} 
c_0 = \frac{2}{\gamma} \left(\frac{r}{H} \right)^2  \, \frac{(\dot{\varpi}_{\rm g}(r_{\rm o})  - \omega)}{\Omega},
\end{equation}
and
\begin{equation} 
c_1 = \frac{2}{\gamma}  \left(\frac{r}{H} \right)^2 \,\frac{1}{\Omega} \, \frac{d \dot{\varpi}_{\rm g}} {dr},
\end{equation}
where $c_1$ and $c_2$ are evaluated at $r=r_{\rm o}$.
Applying equations (\ref{iBC}) and (\ref{norm}), 
we obtain the solution to equation (\ref{Ez}) as
\begin{equation}
E(r)= \frac{Ai(-c_2 - (r-r_{\rm o})/w_e)}{Ai(-c_2)},
\label{Ef}
\end{equation}
where  $Ai$ is the Airy function
and 
\begin{equation}
w_e^3 = \frac{\gamma}{2} \left(\frac{H}{r}\right)^2 \Omega
 \left(\frac{d \dot{\varpi}_{\rm g}}{d  r} \right)^{-1} r_{\rm o}^2 .
 \label{wf}
\end{equation}
Positive constant $c_2$ satisfies 
\begin{equation}
Ai'(-c_2)=0,
\label{Ai'0}
\end{equation}
in order that the outer boundary condition (\ref{oBC}) is satisfied.
There are many (formally infinitely many) values of $c_2$ that satisfy  equation (\ref{Ai'0}), since $Ai$ is oscillatory.
Each such solution corresponds to a free eccentric mode. The number of nodes
increases with increasing values of $c_2$. The fundamental mode has the smallest value of $c_2$ and
for this mode $c_2 \approx 1.019$.

The precession rate is given by
\begin{equation}
\omega = \dot{\varpi}_{\rm g}(r_{\rm o})   - c_{2}\,  w_e \, \frac{d \dot{\varpi}_{\rm g}}{d r}
\label{omf}
\end{equation}
or
\begin{equation}
\omega \simeq \dot{\varpi}_{\rm g}(r_{\rm o}  - c_{2}  w_e).
\label{omfa}
\end{equation}
Therefore, the precession rate for a cool disc with pressure is equal 
to the gravitational precession rate evaluated at a distance $c_2 w_e \simeq w_e $
inside the disc edge.

\begin{figure}
\includegraphics[width=84mm]{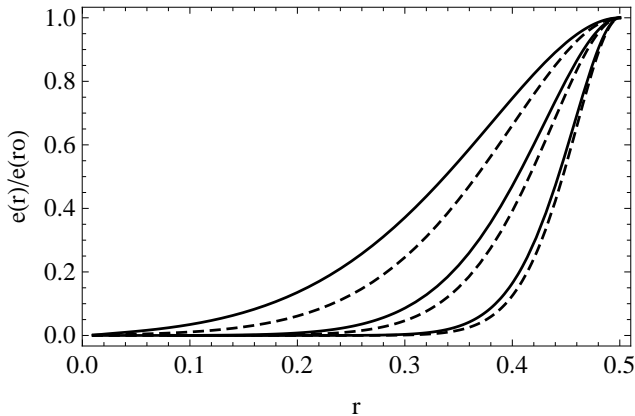}
\caption{Eccentricity distributions for the fundamental free eccentric mode
for systems with standard parameters, but different values
of the dimensionless disc thickness $H/r$ at the disc outer edge. The solid curves are determined by
numerical solutions to equations (\ref{E_eq}) and (\ref{Eom}), and the dashed curves are the
analytic solutions for a cool disc given by equation (\ref{Ef}). The lowest
to highest  dashed and solid curve pairs have values of $H/r=0.005, 0.01,$ and 0.02,
respectively.
         }
\label{fig:ef}
\end{figure}

\begin{figure}
\includegraphics[width=84mm]{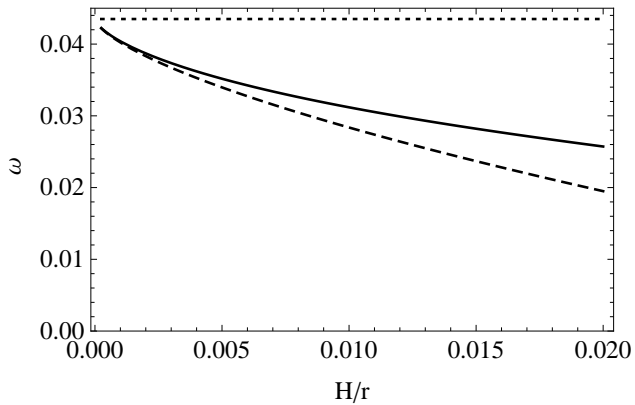}
\caption{Eccentricity precession rate in units of $\Omega_{\rm b}$ plotted against
 $H/r$, the disc thickness-to-radius ratio at the disc outer edge, for the standard model.
The dashed line is given by the analytic formula, equation (\ref{omf}) for a cool disc,
the solid line is determined by numerical solutions to equation (\ref{E_eq}), and the
dotted line is the precession rate of a free particle at the disc edge, $\dot{\varpi}_{\rm g}(r_{\rm o})$. 
         }
\label{fig:omf}
\end{figure}

Equation (\ref{wf}) for width $w_e$ demonstrates the competition between the
sound speed or $H$ that broadens the eccentricity distribution, and differential gravitational  precession,
$d \dot{\varpi}_{\rm g}/dr$, that acts to confine it.  The analytic eccentricity distribution given by equation (\ref{Ef}) 
roughly agrees with the numerically determined distribution for $H/r = 0.02$, but becomes
fairly accurate for $H/r=0.005$ (see Fig.~\ref{fig:ef}). The precession rates
given by equation (\ref{omf}) obtain similar levels of agreement with numerically
determined values (see Fig.~\ref{fig:omf}).

\subsection{Variations in the disc outer radius}

\begin{figure}
\includegraphics[width=84mm]{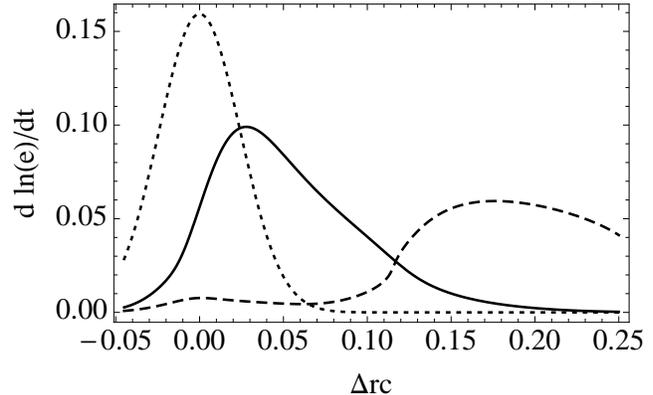}
\caption{
Eccentricity growth rate due to the 3:1 resonance
in units of $\Omega_{\rm b}$ plotted against
 $\Delta r_{\rm c}$ the difference between the disc outer radius and resonance radius. 
In the standard model $\Delta r_{\rm c} \simeq 0.034.$ The solid line is based on smooth continuation of
 the superhump mode. The dashed line is the growth rate for a higher order mode. The dotted
curve is the 3:1 resonance instability factor $s(r_{\rm res} + \Delta r_{\rm c})$ of
equation (\ref{E_eq}) in units of $\Omega_{\rm b}$. Notice that for $\Delta r_{\rm c} \ga 0.12$
this alternative mode dominates, since it has a higher growth rate than the superhump  
 mode.   }
\label{fig:grdr}
\end{figure}

We analyse the eccentricity growth rate due to the 3:1 resonance
as a function of  the disc outer radius, or 
equivalently the clearance between the resonance and disc outer edge $\Delta r_{\rm c}=r_{\rm o}-r_{\rm res}$,
with all other parameters obtained from the standard model.
For larger discs, other resonances can play a role, such as the 2:1 at $r=0.61$ or $\Delta r_{\rm c} = 0.15$,
but we ignore such effects.
The solid curve in Fig.~\ref{fig:grdr} plots the eccentricity growth rate 
for the superhump mode as a function of $\Delta r_{\rm c}$.  If the disc resides
 sufficiently far inside the resonant radius,
that is for $\Delta r_{\rm c} < -w_{\rm res} \sim -0.07$, then there is little overlap between the disc and the
resonant region (i.e., where $s(r)$ is substantial; see equation (\ref{gr2})). The growth rates under such conditions
are quite small. The growth rate rises rapidly for the
disc edge close to, but inside the resonant radius (the peak of the dotted line), that
is for $-0.05 < \Delta r_{\rm c}<0$. Consequently, the growth rate can be quite
sensitive to the disc edge location.

 We have idealised the gas distribution
to be a smooth function of position that is sharply truncated at the disc edge.
In a more realistic description, the disc density distribution  tapers near the disc edge.
The degree of overlap between the disc density distribution can be delicate in the
case of superhump binaries. The reason is that the disc is truncated by tidal forces at a radius
that is close to the 3:1 resonance.
Furthermore, the extent of overlap with the resonant region will likely depend on the disc
turbulent viscosity. We would qualitatively expect in the superhump binary case
that the overlap, and consequently the eccentricity growth rate, 
would increase as the disc viscosity increases, since the disc tendency to spread
outward is increased. Just that type of behavior was
found in the simulations by Kley el al (2008).

\begin{figure}
\includegraphics[width=84mm]{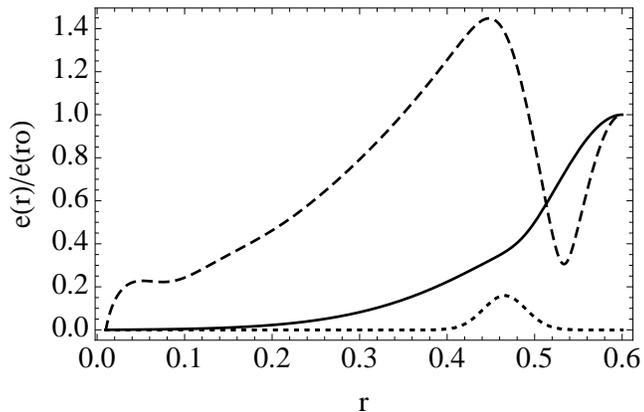}
\caption{Eccentricity distributions for a system with standard parameters, but having
disc outer radius $r_{\rm o}=0.6$ or $\Delta r_{\rm c} \simeq 0.14$. The solid line plots the superhump mode,
while the dashed line plots a higher order mode, corresponding to the rates
plotted in solid and dashed curves respectively in Fig.~\ref{fig:grdr}. The dotted
curve is the 3:1 resonance instability factor $s(r)$ of
equation (\ref{E_eq}) in units of $\Omega_{\rm b}$.
  In this case, the  higher order
  mode  overlaps better with the instability factor $s(r)$ than the superhump mode and consequently
  has a faster growth rate, as seen in Fig.~\ref{fig:grdr}.      }
\label{fig:edr2}
\end{figure}

 For the standard model, $\Delta r_{\rm c} \simeq 0.034$ and the growth rate
is nearly optimal. 
For $\Delta r_{\rm c} \ga 0.05$ the growth rate drops because the peak
amplitude of the mode occurs near the disc edge and is therefore further away from the resonant
region (as seen in equation (\ref{gr2})). This can be seen in comparing for example the solid lines in Fig.~\ref{fig:freem} 
 and Fig.~\ref{fig:edr2} (that corresponds to $\Delta r_{\rm c} \simeq 0.14$).
The former has a stronger overlap with the resonance than the latter.  So the growth rate in the standard disc
case ($\Delta r_{\rm c} \simeq 0.034$) is higher.

For  $\Delta r_{\rm c} \ga 0.12$, the superhump mode
is no longer the fastest growing. Such a large disc is unlikely to occur in typical
superhump binaries, although it could occur in more extreme mass ratio systems. Under such a situation, the growth rates plotted by the
dashed line in Fig.~\ref{fig:grdr} dominate. The dashed line corresponds to a higher order 
eccentric mode. This mode (actually $e^2(r)$, see equation (\ref{gr2})) has a better overlap with $s(r)$,
as seen in Fig.~\ref{fig:edr2}, and consequently grows faster than the superhump mode.
In the absence of a companion, this higher order mode would decay faster in the presence of viscosity
than the 
superhump mode would, due to its smaller scale spatial variations. Other modes could be expected
to dominate under different conditions.

\subsection{Variations in disc sound speed for $0.01 < h < 0.1$}
\label{sec:hvar}

\begin{figure}
\centering%
\resizebox{\linewidth}{!}{%
\plotone{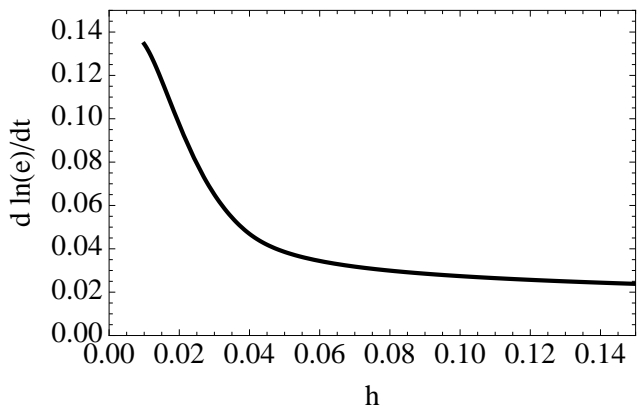}
}
\caption{
Eccentricity growth rate in units of $\Omega_{\rm b}$ plotted against
 $h$, the disc thickness-to-radius ratio at the 3:1 resonance, for the standard model.
   The growth rate is higher at lower $h$ because the eccentric mode becomes more
   confined near the resonance, as seen in Fig~\ref{fig:eH}.      }
\label{fig:grH}
\end{figure}

\begin{figure}
\includegraphics[width=84mm]{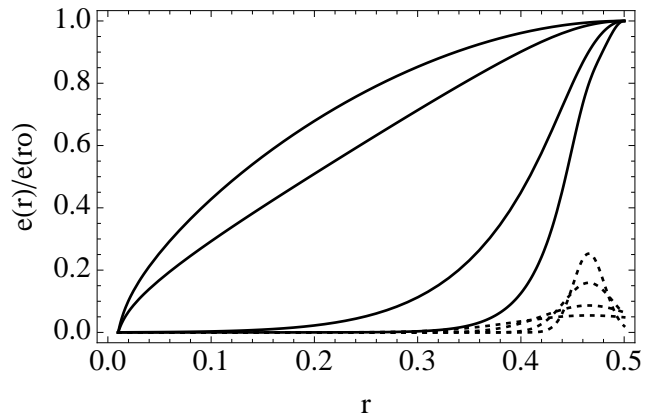}
\caption{Eccentricity distributions for cases with standard parameters, but having
different values of dimensionless disc sound speed $h$, the disc thickness-to-radius ratio at the 3:1 resonance. 
The values of $h$ from the  lowest to highest solid curves are 
0.01, 0.02, 0.05, and 0.1, respectively. The dotted curves are the  3:1 resonance instability factors $s(r)$ of
equation (\ref{E_eq}) in units of $\Omega_{\rm b}$ for the same set of $h$ values, where the
peak values decrease with increasing $h$.
         }
\label{fig:eH}
\end{figure}

We consider variations from the standard model in the disc gas sound speed or equivalently $h$, leaving all other parameters fixed.
Fig.~\ref{fig:grH} shows that the eccentricity growth rate decreases  
with the dimensionless disc sound speed, as measured by dimensionless disc thickness $h$.
Such behavior was found in simulations by Kley et al (2008).
As seen in Fig.~\ref{fig:eH}, this result can be understood in terms of the effects of the sound speed
on mode confinement. As the disc sound speed increases, the mode spreads
further until for $h \sim 0.05$ the eccentricity is fairly uniform. The spreading is not
simply a consequence of the broadening of the instability factor $s(r)$ with $h$ (see equation (\ref{w})), since it does
not spread as rapidly as the mode does with increasing $h$. 
Notice that for large $h$, the growth rate in Fig.~\ref{fig:grH} approaches a constant value.
This behavior can be understood by the fact that in warmer discs the disc eccentricity becomes uniformly 
distributed in radius over the entire disc. We can then crudely take $w_e \sim r_{res}$ and $e_{\rm res}
\sim e_{\rm max}$ in equation (\ref{gr2}) and
obtain a growth rate of $0.02 \Omega_{\rm b}$ that is in rough agreement with the figure for $h \sim 0.1$.

\subsection{Variations in disc sound speed for $ h < 0.01$}

\begin{figure}
\includegraphics[width=84mm]{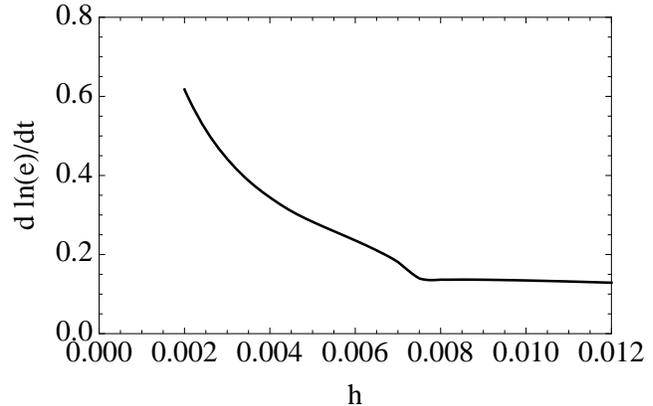}
\caption{
 Eccentricity growth rate in units of $\Omega_{\rm b}$ plotted against
 $h$, the disc thickness-to-radius ratio at the 3:1 resonance, for the standard model.
 This plot extends Fig.~\ref{fig:grH} to colder discs. For $h \la 0.007$, the growth
 rate increases more rapidly with decreasing $h$, as the eccentric mode detaches
 from the disc outer edge and becomes centered on the resonance.
      }
\label{fig:grc}
\end{figure}

\begin{figure}
\includegraphics[width=84mm]{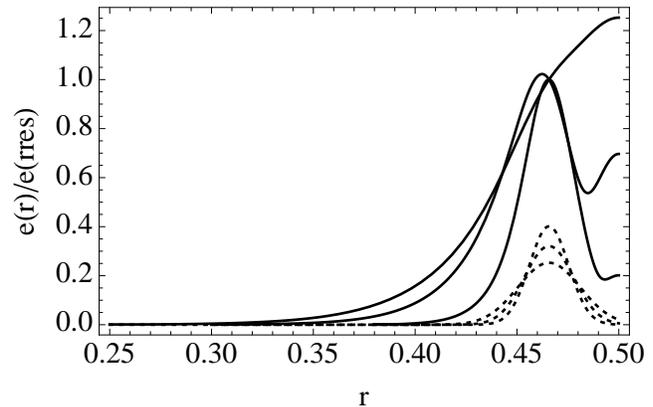}
\caption{Eccentricity distributions, normalised by $e(r_{\rm res})$,
 for cases with standard parameters, but having
different values of dimensionless disc sound speed $h$, the disc thickness-to-radius ratio at the 3:1 resonance. 
The values of $h$ from the  lowest to highest solid curves at $r=0.5$ are 
0.005, 0.007, and 0.01, respectively. The dotted curves are the  3:1 resonance instability factors $s(r)$ of
equation (\ref{E_eq}) in units of $\Omega_{\rm b}$ for the same set of $h$ values, where the
peak values decrease with increasing $h$.  }
\label{fig:ec}
\end{figure}

We analyse the behavior of the eccentricity growth rates and eccentricity
distributions for cooler discs than were considered in Section \ref{sec:hvar}.
Fig.~\ref{fig:grc} shows that the growth rate  rises abruptly with decreasing $h$, for
$h \la 0.007$.  Fig.~\ref{fig:ec} shows that the nature of the eccentricity distributions change
for sufficiently cool discs, since they detach from the disc outer edge.
In the limit of a very cool disc, the resonance width $w_{\rm res} = h^{2/3} r_{\rm res}$ becomes smaller than 
$\Delta r_{\rm c}$, the distance from the resonance to the disc edge. 
The distributions are then centered on the resonance.
 This description is valid
provided that the resonance lies inside the disc, $r_{\rm res} < r_{\rm o}$.
The rapid growth rates for cool discs are a consequence of the narrowing of the width
of the eccentricity distribution (see equation (\ref{gr2})). 

If $r_{\rm res} > r_{\rm o}$, the growth rate will drop to zero with decreasing sound speed, since there will be little
overlap between the disc and the resonance. We assume in the remainder of this
subsection that the resonance does lie within the disc, as follows from the parameters
of the standard model.

We develop an analytic model in the very cool disc limit.
In that limit, we have that $|dE/dr| >> |E|/r$ and consequently equation (\ref{E_eq})
reduces in lowest order to
\begin{equation}
i \Omega( r_{\rm res}) \frac{d^2 E}{d x^2} + s(x) E(x) = i \omega E(x),
\label{E_eqc}
\end{equation}
where $x= \sqrt{2/\gamma} (r - r_{\rm res})/H( r_{\rm res} )$.
The boundary conditions for $E(x)$ are 
\begin{equation}
E(-\infty)= E'(\infty)=0
\label{BCx}
\end{equation}
and the normalization condition is chosen to be
\begin{equation}
E(0)= 1.
\end{equation}

For a cool disc such that $H \ll w_{\rm res}$,  we approximate
 the gaussian representation of $s(x)$ given by equation (\ref{s}) as
\begin{equation}
s(x) \simeq \frac{\chi}{\sqrt{2 \pi} w_{\rm res}} \left(1 - \frac{\gamma}{2} \left(\frac{H}{w_{\rm res}}\right)^2 x^2\right).
\end{equation}
The solution is given by
\begin{equation}
E(r) = \exp{\left[-(1-i) \left(\frac{r-r_{\rm res}}{w_e}\right)^2 \right]}
\end{equation}
and
\begin{equation}
\omega = -\frac{i \chi}{\sqrt{2 \pi} w_{\rm res}} \left(1 -\frac{(1+i)}{4} \left( \frac{w_e}{w_{\rm res}}\right)^2 \right),
\label{omgc}
\end{equation}
where
\begin{equation}
w_e = 2^{5/8} \gamma^{1/4} \pi^{1/8} \sqrt{ Hw_{\rm res}}  \left(\frac{\Omega \, w_{\rm res}}{\chi}\right)^{1/4}.
\label{dregc}
\end{equation}

The eccentricity $e(r)=|E(r)|$ is of gaussian form that is centered on the resonance and $\varpi(r) = (r-r_{\rm res})^2/w_e^2$. The real
radial velocity determined by equation (\ref{u}) is \\$ u'(r,\phi) = -r \Omega \exp{(-(r-r_{\rm res})^2/w_e^2)} 
\sin{((r-r_{\rm res})^2/w_e^2 - \phi)}$.  
For $r <  r_{\rm res}$, the phase term $(r-r_{\rm res})^2/w_e^2 - \phi$ is constant for increasing $r$
and decreasing
$\phi$.  The waveform is then trailing for $r < r_{\rm res}$ and leading for   $r > r_{\rm res}$. 
The width
$w_e$ involves a competition between the resonant growth factor $\chi$ and the
gas sound speed, through its dependence on $H$ and $w_{\rm res}$. For $w_{\rm res}$ given by equation (\ref{w}), it follows that 
$w_e/w_{\rm res}$ goes to zero in the limit that the gas sound speed (or $H$) goes to zero.
That is, as the gas sound speed goes to zero, eccentricity distribution becomes concentrated into a region
that is much smaller than the width of the resonance $w_{\rm res}$.
From equation  (\ref{omgc}), we see that eccentricity growth rate, $-Im(\omega)$, approaches $s(r_{\rm res})$
for a cool enough disc that $w_e \ll w_{\rm res}$.  That is, the eccentricity growth rate approaches the
growth rate
of the instability at the resonance, as is consistent with equation (\ref{gr2}).

Curiously, the detailed properties of the eccentricity distribution depend on the form taken for
$s(x)$. If instead of a gaussian, we adopt a 'top-hat' form, then the results are somewhat different.
Consider $s(x)$ defined by
\begin{equation}
s(x) = \frac{\chi}{2 w_{\rm res} }
\end{equation}
for $  \sqrt{\gamma/2} \, |x| H = |r - r_{\rm res}| < w_{\rm res}$ and zero otherwise.
The solution to equation (\ref{E_eqc}) is then given by
\begin{equation}
E(r) = \cos{\left[ \frac{\pi (r-r_{\rm res})}{2 w_{\rm res}} \right]}
\end{equation}
for  $|r - r_{\rm res}| < w_{\rm res}$ and zero otherwise,
and
\begin{equation}
\omega = -\frac{i \chi}{2 w_{\rm res}}  - \frac{\pi^2 \gamma H^2 \Omega}{8 w_{\rm res}^2 }.
\label{omgh}
\end{equation}
Unlike the gaussian case, the width of the eccentricity distribution does not get smaller
than the width of the resonance in the very cool disc limit.  In this case we have $\varpi(r)=0$,
unlike the gaussian case.
While the gaussian case involved wrapped modes that were leading (trailing)
for $r> r_{\rm res}$ ($r < r_{\rm res}$), the top-hat case contains equal amounts of leading and trailing
waves at each radius.
The growth rate  is $-Im(\omega) = \chi/(2 w_{\rm res}) = s(r_{\rm res})$. So again
 the eccentricity growth rate approaches the growth rate
of the instability at the resonance.


\subsection{Variations of the inner boundary}
 \label{sec:iBC}
 
\begin{figure}
\includegraphics[width=84mm]{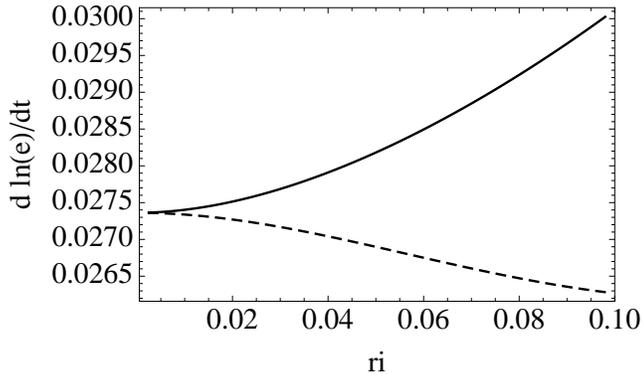}
\caption{Eccentricity growth rate in units of $\Omega_{\rm b}$ plotted against
disc inner radius $r_{\rm i}$ for the standard model with $h=0.1$. The 
solid (dashed) line is for the $E=0$ ( $\partial_r E=0$) inner boundary condition.
The growth rates converge as the disc inner radius goes to zero.
         }
\label{fig:gr-ri}
\end{figure}

We consider here the influence of both the location of the inner boundary and the form
of the inner boundary condition on the eccentricity distribution and growth rate.
For the cases described above with the standard model or colder discs, the inner boundary 
location typically has little influence. In such cases,
 the eccentricity distribution smoothly approaches zero with decreasing $r$ far
from the inner boundary. But for warmer discs, as may occur in the protostellar context,
the eccentricity distribution is nonzero close to the inner boundary, as seen in Fig.~\ref{fig:eH}.
We consider then the case of a standard model disc having $h=0.1$.
The solid line plotted in Fig.~\ref{fig:gr-ri} shows that the growth rate, even in this case,
is insensitive to the inner boundary radius, provided that it is reasonably small, 
$r_{\rm i} \la 0.1$.

In the protostellar disc context, the disc may not extend to the central object,
since there may be a hole near the disc center. 
It is then
not obvious that the hard wall boundary condition (\ref{iBC}) should apply.
We consider then the effects of applying the inner boundary condition $\partial_r E=0$, 
the same boundary condition we applied at the outer boundary. 
To explore these effects, we consider an expansion of equation (\ref{E_eq})
for small $r$ with the standard model. We take $r$ to be dimensionless, normalised
by the binary separation. We apply equation (\ref{Eom}) in the small $r$ limit and obtain 
\begin{equation}
r^2  E''(r)  + 1.5 r   E'(r) - 0.9 E(r)=0.
\end{equation}
The solution is written as
\begin{equation}
E(r) = c_1 \left( r^{s_1} + c_2 r^{-s_2} \right),
\end{equation}
where
$s_1 \approx 0.731$ and $s_2 \approx 1.23$. Constants $c_1$ and $c_2$
are determined by the inner boundary condition and the requirement that
the solution join onto the solution outside of this region of small $r$.
The hard wall ($E=0$) boundary condition requires $c_2=-r_{\rm i}^{s_1+s_2}$.
In the limit that $r_{\rm i}$ goes to zero, we have that $E \propto r^{s_1}$.

\begin{figure}
\includegraphics[width=84mm]{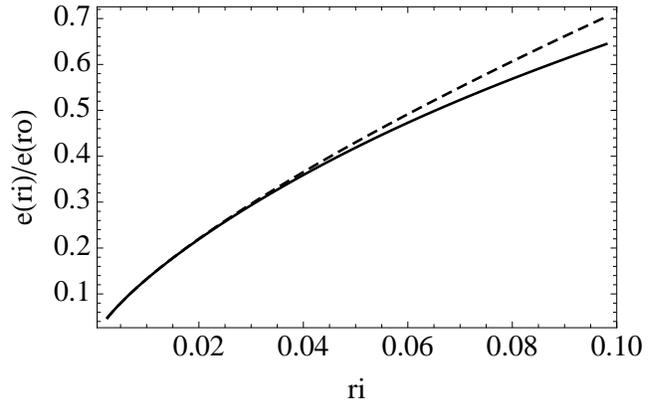}
\caption{Eccentricity at the inner disc edge at a function of the inner disc radius $r_{\rm i}$ 
for the standard model with $h=0.1$. The solid line is the result of numerical
solutions for equations (\ref{E_eq}) and (\ref{Eom}) with the condition
$\partial_r E=0$ applied at both the inner and outer boundaries.
The dashed line is the prediction for $r_{\rm i} \ll 1$ based on equation (\ref{eri})
with $c_3=3.82$.
         }
\label{fig:eri}
\end{figure}

The inner boundary condition  $\partial_r E=0$
requires that $c_2 = s_1 r_{\rm i}^{s_1+s_2}/s_2$.
We expect $c_1$ to be of order unity, since $|E| \sim 1$ where  $r$ of order unity.
We then have that
\begin{equation}
E(r_{\rm i}) = c_3 {r_{\rm i}}^{s_1},
\label{eri}
\end{equation}
for $r_{\rm i}$ small,
where $c_3$ is a constant of order unity.
Therefore, we have that the solution with the  $\partial_r E=0$  inner boundary condition
approaches the solution with the hard wall $E=0$ inner boundary condition in the limit
that $r_{\rm i}$ goes to zero. That is, both cases have the same $E(r_{\rm i})=0$ value in
this limit. Consequently, the  eccentricity distributions and growth rates are the same for both boundary conditions
in the limit that the inner radius goes to zero.
Fig.~\ref{fig:gr-ri} shows that the growth rates are nearly the same for the two
boundary conditions with small $r_{\rm i}$ and that they match as $r_{\rm i}$ goes zero.
Fig.~\ref{fig:eri} verifies that equation (\ref{eri}) holds well for numerical solutions of eccentric modes
when $r_{\rm i}$ is small.

\subsection{Variations of the resonance width}
\label{sec:grw}

\begin{figure}
\includegraphics[width=84mm]{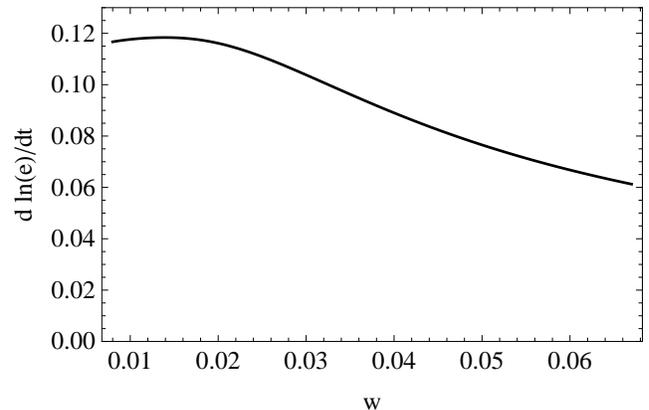}
\caption{Eccentricity growth rate in units of $\Omega_{\rm b}$ plotted against
resonance width $w$ with all other parameters given by the standard model. The growth rate declines somewhat
for larger widths, as the resonance instability distribution $s(r)$ spreads beyond the disc outer edge.
         }
\label{fig:gr-w}
\end{figure}

For the standard model, the resonance width given by equation (\ref{w})  
evaluates
to $w_{\rm res} \simeq  0.034$. Fig.~\ref{fig:gr-w} shows how the eccentricity growth rate varies with $w_{\rm res}$,
where all other parameters are 
fixed at their standard values. For small $w_{\rm res}$, the growth rate approaches a constant
value as the resonance instability function $s(r)$ approaches the form of a Dirac delta function, as was used
by GO06. The properties of the eccentricity distribution are insensitive to the resonance width.
 Recall that for the standard model, the resonance is located at distance
$\Delta r_{\rm c} \simeq  0.034 \simeq w_{\rm res}$ from the disc outer edge. 
The portion of the resonance distribution $s(r)$ that lies outside the disc
does not contribute to eccentricity growth. Consequently, the growth rate declines for
$w_{\rm res} \ga \Delta r_{\rm c}$.

\section{Summary and Discussion}
\label{sec:diss}

We have investigated the evolution of discs subject to the eccentric instability caused by the 3:1
resonance.  
The 3:1 resonance is believed to be responsible for the eccentricity
of discs in superhump binaries and the excitation mechanism is understood in terms
of a mode-coupling model (L91).  
The eccentricity growth rate is affected by the disc eccentricity distribution, principally
by the radial width of the eccentric region and by the eccentricity at the resonance (see equation (\ref{gr2})).
We explored the disc eccentricity distribution and its effect on the growth rate, as a function of system
parameters using a linear equation developed by GO06, based on a two-dimensional
disc model. The effects of turbulent viscosity on the eccentricity were ignored.

For typical superhump binary system parameters, the disc settles into a mode,
the superhump mode, in which the eccentricity is concentrated in the outer parts
of the disc (see Figs.~\ref{fig:mols} and \ref{fig:freem}). The mode results from the excitation of
fundamental free eccentric mode. For the standard model defined in Section \ref{sec:std}, the 
eccentricity growth rate is robust, since the mode overlaps
well with the resonance and is concentrated near it.  If the disc
is somewhat smaller or larger than for the standard model, then the growth rate of the superhump mode is 
lower (Fig.~\ref{fig:grdr}).
Similarly, if tidal forces lower the disc density in the region near the resonance, then we expect
the growth rate to be reduced. In addition, the disc tapering needed to avoid Rayleigh instability at the outer
edge could cause some density reduction in the resonant region and lower the growth rate.

The width of the eccentricity distribution
is controlled by a competition between pressure forces that act to spread the
mode and  the eccentricity growth and differential precession that act to confine it (Figs.~\ref{fig:e-q1}
and \ref{fig:eH}). 
The growth rate
decreases with the increasing disc sound speed or disc thickness (see Fig.~\ref{fig:grH})
and increases with binary mass ratio somewhat faster than quadratically
(Fig.~\ref{fig:grq}). 
Other eccentric modes can be excited under different conditions, such as larger
or colder discs (Figs.~\ref{fig:edr2} and \ref{fig:ec}).

Once the superhump feature is observed in binary systems, it is likely that the exponential
growth phase, which occurs for small eccentricity, has ceased. 
In that case, the disc eccentricity may then be best described by
the fundamental free eccentric mode, the free mode excited by the superhump instability.
Equation (\ref{omf}) describes the free precession rate of a cool disc, subject to
the approximations of this model.

Discs in young binary star systems of extreme mass ratio can similarly undergo eccentric
instability. The eccentricity growth rates in units of the binary frequency are likely lower than
in the superhump binary case, since protostellar discs are considerably warmer as
measured by the dimenionless thickness $h$  (see Fig.~\ref{fig:grH}).  Star-planet
systems could similarly undergo such instability, although it likely involves the participation of other
resonances that lie closer to the planet (Kley \& Dirksen 2005, D'Angelo et al 2006).
In that case also, we expect the eccentricity of the inner disc to be fairly broadly
distributed because the mass ratio is much smaller and the disc thickness is
larger than in the superhump binary case.

The effects of the disc vertical thickness, turbulent viscosity, and nonlinearity, omitted in this study,
should be explored in the future (e.g., Ogilvie 2001).

\section{Acknowledgments}

I have greatly benefitted from many discussions with Gordon Ogilvie and Jim Pringle on this topic.
I am grateful for support from the IoA visitor program and
  NASA 
grant NNX07AI72G.


\end{document}